\documentclass[10pt,letterpaper,twocolumn]{article}
\usepackage{ol2}
\usepackage[draft]{hyperref}

\usepackage{graphicx}
\usepackage{eurosym}
\usepackage{amsfonts}
\usepackage{amsmath}
\usepackage{amsopn}
\usepackage{amssymb}
\usepackage{graphicx}
\usepackage{graphics}
\usepackage{color}
\usepackage{verbatim}

\newcommand {\lan}{\langle}
\newcommand {\ran}{\rangle}
\newcommand {\sG}{\sigma_{\Gamma}}
\newcommand {\sK}{\sigma_{K}}
\newcommand{\PT}{{\cal PT}}

\begin{document}

\twocolumn[

\title{Stochastic $\PT$-symmetric coupler}

\author{V. V. Konotop and D. A. Zezyulin}

%\affiliation{
\address{
 Centro de F\'isica Te\'{o}rica e Computacional and Departamento de
F\'isica, Faculdade de Ci\^encias, Universidade de Lisboa, Avenida Professor
Gama Pinto 2, Lisboa 1649-003, Portugal
}
\date{\today}

\begin{abstract}
We introduce a stochastic $\PT$-symmetric coupler, which is based on dual-core   waveguides with fluctuating parameters, such that the gain and the losses are  exactly balanced  in average.
%
%
%in average the condition of the exactly balanced gain and losses are satisfied. 
We consider different parametric regimes which correspond to the broken and unbroken $\PT$ symmetry, as well as to the exceptional point of the underlying deterministic system. We demonstrate that in all the cases the statistically averaged intensity  of the field grows.
This result holds for either  linear or nonlinear coupler and   is independent on the type of fluctuations.
%The growth occurs in the active waveguide and depending on the system parameters %may also occur or not in the absorbing waveguide. The evolution of the mean fields %is described by the passive $\PT$-symmetric system, with fluctuations of the %gain/loss and of the coupling acting as effective gain and dissipation, %respectively.
\end{abstract}


]

%\pacs{}
%\maketitle

Parity-time ($\PT$) symmetry can be viewed as a property of a physical system to possess delicate balance between the dissipation and the gain~\cite{Bender}. In optics this  property is expressed by the constraint imposed on the refractive index, $n(x)=n^*(-x)$~\cite{Muga}.  If the gain and the losses are not {\em exactly} balanced, the system becomes dissipative or active. One however still can observe effects related to $\PT$ symmetry provided the system is linear. This occurs in the so-called passive $\PT$-systems because a qualitative change in the field evolution (i.e. the $\PT$ symmetry phase transition) takes place when the system parameters cross an exceptional point (see e.g.~\cite{Heiss}). In such a setting the $\PT$ symmetry phase transition was detected experimentally for the first time~\cite{Guo}. Mathematically, the existence of linear passive $\PT$-symmetric devices is explained by the possibility of scaling out the average decay or gain by a proper exponential factor, thus reducing the description to an effective system with properly matched and balanced gain and dissipation. The simplest geometry allowing one to ensure this last requirement is a two-waveguide structure~\cite{Guo2}, which is used in most experimental setups~\cite{Ruter,exp1}.

Scaling out dissipation (gain) is impossible for a  nonlinear system. To observe nonlinear $\PT$-related  phenomena it is crucial to have a  system with  real gain and absorption which are balanced. A linear system obeying the latter   requirement  was experimentally created in~\cite{Ruter} using doped waveguides. Alternative possibilities, like usage of plasmonics~\cite{Lupu} or hollow core waveguides filled with resonant multilevel atomic gasses~\cite{HHK,CZKH}, were  discussed theoretically.

In any experimental setting, like in the optical ones mentioned above, in electrical~\cite{Kottos} or in mechanical~\cite{mechanical} systems, the $\PT$-symmetric distribution of the parameters can be viewed only as an averaged effect, perturbed by presence of imperfections of the structure or thermal fluctuations. Even in high precision waveguides the random fluctuation can
destroy the delicate balance between the absorption and the gain. This rises a question about persistence of the properties of a deterministic $\PT$-symmetric system when its parameters are subject to small random variations.

In this Letter we address   behavior of a random (linear or nonlinear) $\PT$-symmetric coupler.
Our stochastic model   accounts for the    random coupling $\kappa + K(z)$,
with $\kappa$ being the mean value of the coupling and $K(z)$ describing random deviations,
and  random deviations $\Gamma_{1,2}(z)$
of the gain in the first waveguide and dissipation in the second waveguide, which in average are balanced and are described by the gain/loss coefficient $\gamma$.
The respective dynamical system which will be referred to as a stochastic $\PT$-symmetric dimer reads (an overdot stands for the derivative with respect to evolution variable $z$):
\begin{eqnarray}
\label{main}
\begin{array}{lcr}
i\dot{q}_{1}=[\kappa+ K(z)]q_2+i[\gamma+\Gamma_1(z)]q_1 +\chi |q_1|^2q_1,
\\
i\dot{q}_{2}=[\kappa + K(z)]q_1-i[\gamma+\Gamma_2(z)]q_2 +\chi |q_2|^2q_2.
\end{array}
\end{eqnarray}
 For $K(z)\equiv 0$ and $\Gamma_{1,2}(z)\equiv 0$  Eqs.~(\ref{main})  are   reduced to the well known (deterministic) nonlinear $\PT$-symmetric coupler, which was introduced in~\cite{Ramezani,Sukhorukov}.

We address  the situation where uncontrollable small deviations in the deterministic
parameters can be approximated by the Gaussian delta-correlated random processes (white noises) with zero mean values. Since in practice physical mechanisms responsible for gain, dissipation and coupling are different, we consider all random processes to be statistically independent. Finally, we assume that the dispersions of the fluctuations of the gain and of the dissipation are equal. All   these suppositions are expressed by the following statistical characteristics:
\begin{eqnarray}
\label{mean}
\langle\Gamma_{1,2}(z)\rangle=\langle K(z)\rangle =
\langle K (z)\Gamma_{1,2}(z')\rangle=0,
\label{linear_stat}
\\
\langle \Gamma_j(z)\Gamma_k(z')\rangle=2\delta_{jk}\sigma_\Gamma^2\delta(z-z'),
\\
\langle K(z)K(z')\rangle=2\sigma_K^2\delta(z-z'),
\end{eqnarray}
where   $\sigma_{K,\Gamma}>0$   characterize dispersions of fluctuations and angular brackets stand for the statistical average.

We are interested in the mean fields $\lan q_{1,2}\ran$ and in the
 correlators $\lan q_jq_k^*\ran$, which at $j=k$ give average intensities of the fields. 
 To address the evolution of the correlators 
we rewrite   (\ref{main}) in terms of the Stokes components~\cite{Eilbeck}:
\begin{eqnarray*}
\label{spin}
\begin{array}{ll}
S_0=|q_1|^2+|q_2|^2, &   S_1=q_1 q_2^*+q_1^*q_2,
\\
S_3=|q_1|^2-|q_2|^2, &
S_2=i(q_1 q_2^*- q_1^*q_2),
\end{array}
\end{eqnarray*}
which satisfy the identity $S_0^2=S_1^2+S_2^2+S_3^2$  and solve the system of stochastic equations
\begin{eqnarray}
\label{S0}
 \dot{S}_0=2\gamma_-(z)  S_0+2[\gamma+\gamma_+(z)] S_3,
\\
\label{S1}
 \dot{S}_1=2\gamma_-(z) S_1-\chi S_2S_3,
\\
\label{S2}
 \dot{S}_2=2\gamma_-(z) S_2-2[\kappa+K(z)] S_3+\chi S_1S_3,
\\
\label{S3}
 \dot{S}_3=2\gamma_-(z) S_3+2[\gamma+\gamma_+(z)] S_0
+2[\kappa+K(z)] S_2,
\end{eqnarray}
where we introduced the random processes
$
\gamma_\pm (z)=\frac12 [\Gamma_1(z)\pm\Gamma_2(z)],
$
which have zero mean values $\lan \gamma_{\pm}\ran=0$  and correlators
\begin{eqnarray}
\label{statistics}
\lan\gamma_\pm(z)\gamma_\pm(z')\ran=\sG^2\delta(z-z'), \,\,\,
\lan\gamma_-(z)\gamma_+(z')\ran =0.
\end{eqnarray}

%\paragraph{Linear coupler --}
Let us start with  the linear problem, letting $\chi=0$  in Eqs.~(\ref{main}).
It is convenient to  split  the fields into     real and imaginary parts which    satisfy   two decoupled systems of   real equations. In particular, introducing
 $ x_1=\mbox{Re}\, q_1$  and  $x_2=\mbox{Im}\, q_2$
we obtain the following system:
% of  stochastic equations:
\begin{eqnarray}
\label{lin_stoch}
\begin{array}{l}
\dot{x}_1=
[\gamma+\Gamma_1(z)]x_1+[\kappa+K(z)]x_2,
\\
\dot{x}_2=-[\gamma+\Gamma_2(z)]x_2-[\kappa+K(z)]x_1.
\end{array}
\end{eqnarray}
To obtain the dynamics of the averaged fields we use the Furutsu-Novikov formula (see e.g.~\cite{Konotop,Klyatskin}) to compute
\begin{eqnarray}
\label{FN_G}
\lan K(z)x_k(z)\ran=\int_{0}^{z}\lan K(z)K(z')\ran \left\langle
\frac{\delta x_k(z)}{\delta K(z')} \right\rangle dz'
\nonumber
\\
=(-1)^{3-k}\sK^2\lan x_{3-k} (z)\ran \qquad (k=1,2)
\end{eqnarray}
and, analogously,  $\lan\Gamma_j(z)x_k(z)\ran =\sG^2\delta_{jk}\lan x_j (z)\ran$.
%and $\lan K(z)x_k(z)\ran=(-1)^{3-k}\sK^2\lan x_{3-k} (z)\ran$ ($j,k=1,2$).
 This gives the following simple equation for the mean values:
\begin{eqnarray*}
\label{M1}
\left(\!\!\!
\begin{array}{c}
\lan \dot{x}_1 \ran
\\
\lan \dot{x}_2 \ran
\end{array}
\!\!\!\right)
=
\left(\!\!
\begin{array}{cc}
 \sG^2-\sK^2+\gamma \!& \kappa
\\
-\kappa &  \sG^2-\sK^2-\gamma
\end{array}
\!\!\right)\left(\!\!\!
\begin{array}{c}
\lan  {x}_1\ran
\\
\lan {x}_2 \ran
\end{array}
\!\!\!\right).
\end{eqnarray*}
Thus fluctuations of the gain/loss and the coupling coefficients introduce effective gain and dissipation, respectively, the combined effect being defined by $\sG^2-\sK^2$, i.e. for the averaged fields we obtained a passive $\PT$-symmetric coupler. By scaling out the induced dissipation: $\lan  x_{1,2} \ran\sim e^{(\sG^2-\sK^2)z}$, one ensures that the exceptional point persists at $\gamma=\kappa$ and determines the  $\PT$ symmetry phase transition. Notice that the obtained result is also valid for different dispersions of the gain and losses: if $\langle \Gamma_j(z)\Gamma_j(z')\rangle=2\sigma_{\Gamma_j}^2\delta(z-z')$ ($j=1,2$) with $\sigma_{\Gamma_1}\neq \sigma_{\Gamma_2}$, then  the system still becomes $\PT$ symmetric after a proper scaling.

\begin{figure}%[htp]
{\includegraphics[width=0.9\columnwidth]{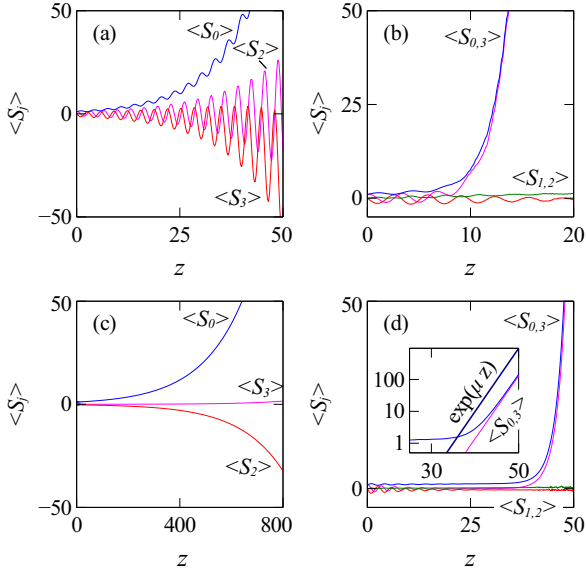}}%
\caption{(Color online) (a) and (c)  Evolution  of the  linear dimer   ($\chi=0$).     (b) and (d)  Evolution of  the nonlinear dimer ($\chi=0.5$). In panels (a) and (b)  $\sigma_\Gamma=0.15$, $\sigma_K=0$, while in panels (c) and (d)  $\sigma_\Gamma=0$, $\sigma_K=0.15$.
In all panels $\gamma=0.25$, $\kappa=1$ and the initial conditions are $S_0(0)=S_3(0)=1$, $S_1(0)=S_2(0)=0$. The linear dynamics are obtained by means of analytical solution of system  (\ref{S0_lin})-(\ref{S3_lin}), while the nonlinear dynamics are obtained numerically, by averaging over   $10^4$ realizations. 
%Blue, green, red and magenta curves correspond to $\lan S_{0, 1,2,3} (z) \ran$, respectively. 
In panels (a) and (c)  $\langle S_1(z) \rangle \equiv 0$ and is not shown. The inset  in   (d) shows a semilogarithmic plot of $\lan S_{0,3}\ran$  together with the straight line which represents the dependence $\propto e^{\mu z}$ with $\mu$ computed from Eq.~(\ref{eq:mu}).
}
\label{fig-unbrok}
\end{figure}

Remarkably, when the fluctuations of the dissipative terms and coupling have the same level of disorder (i.e. $\sG=\sK$) the mean field  is governed by the exactly $\PT$-symmetric system. This however does not mean  that the $\PT$ symmetry is fully reintroduced, which is easily seen
from the analysis of the average values of the Stokes components $S_j$.
Indeed, in the linear case the equation for $\langle S_1\rangle$ is singled out and  gives
the integral ($\dot{J}_{lin} \equiv0$)
\begin{eqnarray}
J_{lin}=(q_1 q_2^*+q_1^*q_2)\exp\left(-2\int_0^z \gamma_-(z')\,dz'\right),
\end{eqnarray}
while the other mean values solve the system
\begin{eqnarray}
\label{S0_lin}
\lan \dot{S}_0\ran=4\sG^2 \lan S_0\ran+2\gamma\lan S_3\ran,
\\
\label{S2_lin}
\lan \dot{S}_2\ran=(2\sG^2-4\sK^2)\lan S_2\ran-2\kappa\lan S_3\ran,
\\
\label{S3_lin}
\lan \dot{S}_3\ran=2\gamma\lan S_0\ran+2\kappa \lan S_2\ran+ 4(\sG^2 -\sK^2)\lan S_3\ran.
\end{eqnarray}
For weak fluctuations,  $\sigma_{\Gamma,K} \ll 1$, from the characteristic equation for this system ($\lan S_j\ran\propto e^{\lambda z}$) we compute with the accuracy $\mathcal{O}(\sG^4+\sG^2\sK^2+\sK^4)$:
\begin{eqnarray*}
\label{lam0}
\lambda_0 =  2\sG^2+2\frac{\sG^2\kappa^2+2\sK^2\gamma^2}{\kappa^2-\gamma^2},
\\
\label{lam12}
\lambda_{\pm}=\pm 2i\sqrt{\kappa^2-\gamma^2}+3\sG^2-2\sK^2
-
\frac{\sG^2\gamma^2 +2\sK^2 \kappa^2}{\kappa^2-\gamma^2}.
\end{eqnarray*}
Thus, even if the $\PT$ symmetry of  the  underlying deterministic problem ($\sG=\sK=0$) is unbroken, i.e. $\kappa>\gamma$,  for the stochastic problem one has  $\lambda_0>0$, i.e. the instability (due to the stochastic parametric resonance~\cite{Klyatskin}) manifesting itself in the exponential growth of the field intensity, takes place.
Two examples of the unstable behavior are shown in Figs.~\ref{fig-unbrok}(a) and (c).  In Fig.~\ref{fig-unbrok}(a) the  instability is  originated by fluctuations of the gain and dissipation ($\sG>0$, $\sK=0$) while in  Fig.~\ref{fig-unbrok}(c) the instability is induced by the randomness of the coupling ($\sG=0$, $\sK>0$).  Figures~\ref{fig-unbrok}(a) and (c) feature   very different   scales in the propagation distance $z$,  which is a result of the difference between the dominating growth exponents [$\lambda_0 \approx 0.093$ in (a) and $\lambda_0 \approx 0.006$ in (c)], and different behaviors with respect to the energy exchange between the waveguides: energy exchange grows in panel (a) ($\textrm{Re}\, \lambda_{\pm}\approx 0.066$) and is attenuated in panel (c) ($\textrm{Re}\, \lambda_{\pm}\approx -0.093$). In these cases, however, we observe that $\lan S_3\ran/\lan S_0\ran\to 0$  along the propagation distance which means that the  energy  grows in the both (i.e. active and absorbing) waveguides.

Turning to   the nonlinear dimer ($\chi>0$), we report evolution of averaged Stokes components  in  Figs.~\ref{fig-unbrok}(b) and (d) which  feature several interesting points.   First, the figures display almost identical growths of  $\lan S_0\ran$ and $\lan S_3\ran$ meaning that almost the whole energy is concentrated in the active waveguide. This corroborates with the fact that a deterministic nonlinear dimer obeys blowing up solutions~\cite{PelinKevr,Barashenkov}, and it is expectable that such (exponentially growing) trajectories dominantly contribute to blow up of the average quantities.
Indeed, random fluctuations ``draw'' a trajectory of the dynamical system  (\ref{S0})-(\ref{S3}) to a domain of instability (reached at some propagation distance) after which the trajectory $\lan |q_1|^2\ran$ starts growing.
For sufficiently large $z$ the mean values   $\lan S_1 \ran$ and $\lan S_2\ran$ become negligibly small compared to $\lan S_0\ran$ and $\lan S_3\ran$. Averaging Eqs.~(\ref{S0}) and (\ref{S3}) and neglecting $\lan S_2 \ran$,  one can estimate  growth rate for $\langle S_{0,3} \rangle$  in the leading order:
\begin{equation*}
\label{eq:S0gr}
\langle S_{0} \rangle  \approx \gamma C e^{\mu z},
\quad
\langle S_{3} \rangle \approx  \left(\sqrt{\sigma_K^4 + \gamma^2} - \sigma_K^2\right) C e^{\mu z},
\end{equation*}
where
\begin{equation}
\label{eq:mu}
\mu=4\sigma_\Gamma^2 - 2\sigma_K^2  + 2\sqrt{\sigma_K^4 + \gamma^2}>0,
\end{equation}
and $C$ is a constant.  The obtained  result  reveals an interesting feature:  if  the  fluctuations of the coupling are present, i.e. $\sigma_K>0$ [as in Fig.~\ref{fig-unbrok}~(d)], then the average energy in the coupler arm with dissipation, i.e.  $2\langle |q_2|^2\rangle=\langle S_{0} \rangle - \langle S_{3} \rangle$, grows exponentially.  If, however, $\sigma_K=0$  [as in Fig.~\ref{fig-unbrok}~(b)], then $\langle S_{0}\rangle$ and $\langle S_{3} \rangle$ grow with the same velocity in the leading order and the behavior of the difference $\langle S_{0} \rangle - \langle S_{3} \rangle$ is determined by the corrections of the next orders. Our numerical results for   Fig.~\ref{fig-unbrok}~(b)  indicate  that $\langle |q_2|^2\rangle$ oscillating slowly decreases (we could not  ensure whether $\lan|q_2|^2\ran$ eventually vanish  at $z\to \infty$;  either way, this issue has no physical relevance because of exponential growth of the intensity in the active arm).

Further we notice that Figs.~\ref{fig-unbrok}(b) and (d) feature comparable scales
for the growth  of $\lan S_{0} \ran$ and $\lan S_{3} \ran$ [contrary to those in Figs.~\ref{fig-unbrok}(a) and (c)].
Besides, the nonlinear  growth of $\lan S_{0,3}\ran$ is more rapid  compared to the linear case. This indicates that the nonlinear resonances are responsible for the observed growth of the intensity. It is interesting that even though  the nonlinearity is responsible for the blow-up dynamics (as discussed above), the growth exponent does not depend of the nonlinearity strength $\chi$ itself.
The inset in Fig.~\ref{fig-unbrok}(d) illustrates that the growth exponent $\mu$ obtained in (\ref{eq:mu}) indeed
matches the curves for $\langle S_{0} \rangle$ and    $\langle S_{3} \rangle$  in   Figs.~\ref{fig-unbrok}(b) and (d).

We also notice  that the nonlinearity suppresses the oscillations of $\lan S_2\ran$ [c.f.  Fig.~\ref{fig-unbrok}~(a)  and  Fig.~\ref{fig-unbrok}(b)].

\begin{figure}%[htp]
{\includegraphics[width=0.9\columnwidth]{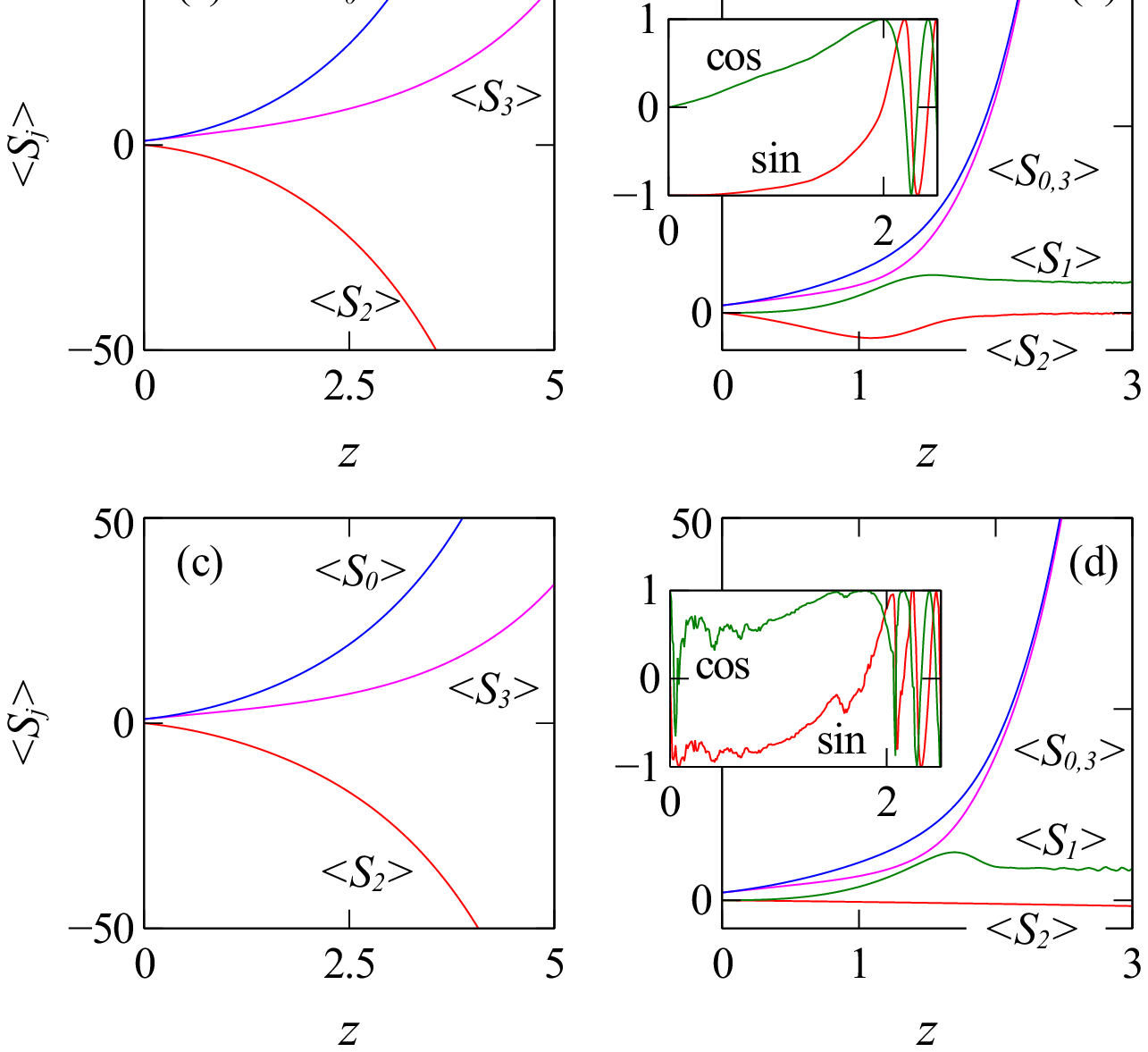}}%
\caption{(Color online) The same as in Fig.~\ref{fig-unbrok} but with  $\gamma=1$ (the exceptional point of the deterministic problem). The insets in   (b) and (d) show dependencies $\lan \cos(\Delta\varphi)\ran $ and $\lan \sin(\Delta\varphi)\ran $
(green and red curves, respectively) for a single arbitrarily chosen realization.}
\label{fig-exp}
\end{figure}

Returning to the linear case, 
we proceed to
the case of the  broken $\PT$ symmetry ($\gamma>\kappa$) as well as for the exceptional point ($\gamma=\kappa$). The evolution of the Stokes components in the latter case is illustrated in Figs.~\ref{fig-exp}(a) and (c). Comparing these two panels we observe that the effects of gain/loss and coupling fluctuations are qualitatively very similar.
The   quantities $\lan S_j\ran$  grow faster than for the unbroken $\PT$ symmetry [Figs.~\ref{fig-unbrok}~(a) and (c)]. Indeed, from (\ref{S0_lin})--(\ref{S3_lin}) one finds that for $\gamma=\kappa$  the growth exponent is given as  $\lambda_0 \approx [8\kappa^2(\sigma_\Gamma^2 +2\sigma_K^2)]^{1/3}$,
 which yields
 $\lambda_0 \approx 0.64$ in Figs.~\ref{fig-exp}~(a) and  $\lambda_0 \approx 0.65$ in Figs.~\ref{fig-exp}~(c).
 Similar to  the case of the  unbroken  $\PT$ symmetry, at $\kappa=\gamma$ we observe exponential growth of the field in the  arm with dissipation. However, in the nonlinear case  [shown in   Figs.~\ref{fig-exp}(b) and (d)]   the value $\langle |q_2|^2 \ran$ grows only if   fluctuations of coupling are present ($\sK>0$). For $\sK=0$ and $\sG>0$ the fluctuations of gain and losses result in decay of  $\langle |q_2|^2\rangle $ as $z\to\infty$.

Another interesting feature observed in Fig.~\ref{fig-exp}~(b)
is that $\lan S_1 \ran\to 4$ and $\lan S_2 \ran\to 0$ as $z\to \infty$. In order to understand this issue, let us first consider only fluctuations of the gain and losses (i.e. $\sK= 0$). From the numerics we conclude that $S_0$ and $S_3$ grow practically in a deterministic way and $\lan S_2\ran \sim\lan\sin(\Delta\varphi)\ran\to 0$,   where $\Delta\varphi$ is a difference between the phases of the fields in the waveguides which is randomized due to the fluctuations (evolutions of the phases for two   realizations are illustrated  in insets in Fig.~\ref{fig-exp}).
Then, after averaging Eq.~(\ref{S2}) one concludes that this is possible only if $\lan S_1\ran \sim\lan \cos(\Delta\varphi)\ran \to 2\kappa/\chi$ (which is  equal to $4$ in all our numerics). Fluctuations of the coupling do   not affect these qualitative arguments because as follows from (\ref{mean}) and almost deterministic behavior of $S_3$, the correlator $\langle K(z) S_3\rangle$ is of order of the fluctuations of the dispersion of $S_3(z)$, which is negligible.

\begin{figure}%[htp]
{\includegraphics[width=0.9\columnwidth]{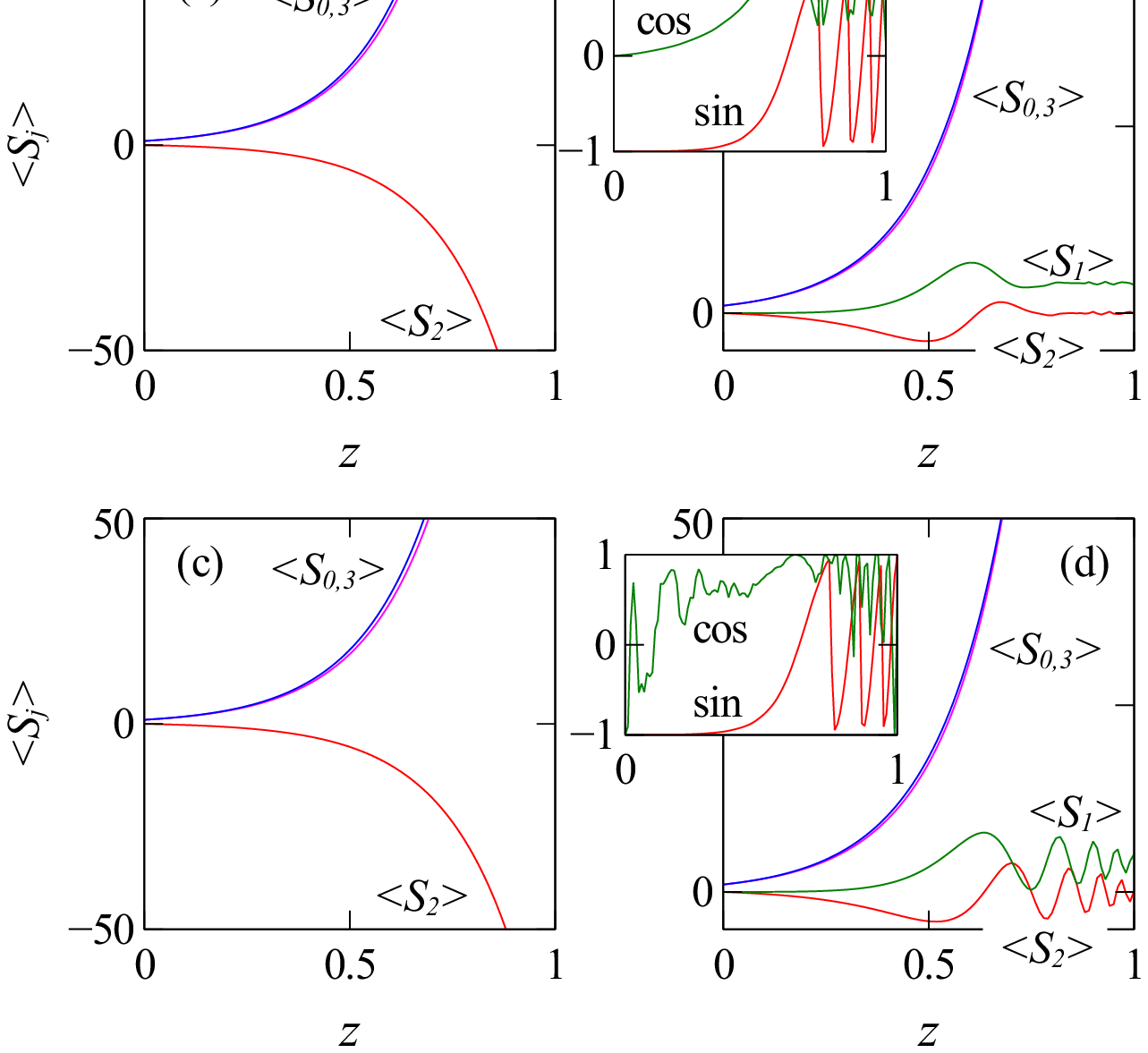}}%
\caption{(Color online) The same as in Fig.~\ref{fig-exp} but with  $\gamma=3$ (the broken $\PT$ symmetry of the deterministic problem).
}
\label{fig-broken}
\end{figure}

We finally  address  the case of the broken   $\PT$ symmetry ($\gamma>\kappa$). A typical outcome for the nonlinear system  is shown in  Fig.~\ref{fig-broken} (b) and (d). Comparing these results with those for the exceptional point [Fig.~\ref{fig-exp}], we observe certain  resemblance, in particular, in quasi-deterministic growth of $S_0$ and $S_3$
and in the asymptotic tendency of 
 $\lan  S_{1,2} \ran$ determined by the phase mismatch of the fields in the waveguides. The main peculiarities of this case are  the growth exponents, which are given by $2\sqrt{\gamma^2- \kappa^2}$ and  practically identical growth of $\lan S_0\ran$ and $\lan S_3\ran$ meaning stronger concentration of the energy in the waveguide with gain. The insets in Fig.~\ref{fig-broken} (b) and (d) display even more regular evolution of the phase mismatch.

All numerical outcomes presented above were obtained for fields  applied to input of the active waveguide. We have also 
studied the cases when the only illuminated  waveguide  is the absorbing one, as well as when the input light is distributed between two waveguides. All the considered initial conditions give statistics very similar to the reported above.

To conclude, we considered  
field propagation in linear and nonlinear stochastic  $\PT$-symmetric coupler. Our main finding is that independently on whether the $\PT$ symmetry of the underlying  deterministic system  is broken or not and independently on the type of fluctuations the statistically averaged intensity  of the field grows. This growth is always observed in the waveguide with gain and depending on the system parameters may be observed or not in the absorbing waveguide (the latter possibility exists for the nonlinear case). The evolution of the mean field preserves the properties of the (passive) $\PT$-symmetric dynamics, where fluctuations of the gain/loss and of the coupling act respectively as the effective gain and dissipation (when they compensate each other the exact $\PT$ symmetry for the system governing the mean field is restored).  As a final remark, in this Letter we left open the effect of fluctuations of the propagation constant 
or the effective Kerr coefficient (the latter stemming from possible imperfectness of the transverse cross-sections of the waveguides). These questions require a separate  study.  

\smallskip

%\section*{ Acknowledgments}
VVK acknowledges discussions with Profs. A. Lupu and H. Benisty.
The work was supported by FCT through the FCT grants PEst-OE/FIS/UI0618/2011 and  PTDC/FIS-OPT/1918/2012.

\pagebreak

\section*{Fifth informational page}

\end{document}